\documentclass[preprint, secnumarabic, amssymb, nobibnotes, aps,a4paper,coloumn,floatfix,showpacs,showkeys,amsmath,altaffilletter,superscriptaddress]{revtex4-1}

\usepackage{graphicx}
\usepackage{pslatex}
\usepackage{xspace}
\usepackage{subfigure}
\usepackage[latin1]{inputenc}
\usepackage[T1]{fontenc}
\usepackage{amsmath}
\usepackage{textcomp}
\usepackage{color}

\usepackage{ulem}

\usepackage{float}

\begin{document}

\title{Anisotropic electron mobility studies on Cl2-NDI single crystals and the role of static and dynamic lattice deformations upon temperature variation}

\author{N.H.~Hansen}
\affiliation{Experimental Physics VI, Universit\"{a}t W\"{u}rzburg, 97074 W\"{u}rzburg, Germany}
\author{F.~May}
\affiliation{InnovationLab GmbH, Speyerer Strasse 4, 69115 Heidelberg, Germany and BASF SE, 67056 Ludwigshafen, Germany}
\author{D.~K\"{a}lblein}
\affiliation{InnovationLab GmbH, Speyerer Strasse 4, 69115 Heidelberg, Germany and BASF SE, 67056 Ludwigshafen, Germany}
\author{T.~Schmeiler}
\affiliation{Experimental Physics VI, Universit\"{a}t W\"{u}rzburg, 97074 W\"{u}rzburg, Germany}
\author{C.~Lennartz}
\affiliation{InnovationLab GmbH, Speyerer Strasse 4, 69115 Heidelberg, Germany and BASF SE, 67056 Ludwigshafen, Germany}
\author{R.~Sanchez-Carrera}
\affiliation{BASF Corporation, 540 White Plains Road, Tarrytown, NY 10591-9005, USA}
\author{A.~Steeger}
\affiliation{Experimental Physics VI, Universit\"{a}t W\"{u}rzburg, 97074 W\"{u}rzburg, Germany}
\author{C.~Burschka}
\affiliation{Institut f\"{u}r Organische Chemie and Center for Nanosystems Chemistry, Universit\"{a}t W\"{u}rzburg, 97074 W\"{u}rzburg, Germany}
\author{M.~Stolte}
\affiliation{Institut f\"{u}r Organische Chemie and Center for Nanosystems Chemistry, Universit\"{a}t W\"{u}rzburg, 97074 W\"{u}rzburg, Germany}
\author{F.~W\"{u}rthner}
\affiliation{Institut f\"{u}r Organische Chemie and Center for Nanosystems Chemistry, Universit\"{a}t W\"{u}rzburg, 97074 W\"{u}rzburg, Germany}
\author{J.~Brill}
\affiliation{InnovationLab GmbH, Speyerer Strasse 4, 69115 Heidelberg, Germany and BASF SE, 67056 Ludwigshafen, Germany}
\author{J.~Pflaum}
\affiliation{Experimental Physics VI, Universit\"{a}t W\"{u}rzburg, 97074 W\"{u}rzburg, Germany}
\affiliation{ZAE Bayern, Magdalene-Schoch-Strasse 3, 97074 W\"{u}rzburg, Germany}

\date{\today}

\begin{abstract} % max 100 words
The anisotropic electron transport in the (001) plane of sublimation-grown Cl$_{2}$-NDI (naphthalene diimide) single crystals is analysed over a temperature range between 175 K and 300 K. Upon cooling from room temperature to 175 K the electron mobility along the direction of preferred transport monotonously increases from 1.5 cm$^{2}$/Vs to 2.8 cm$^{2}$/Vs according to a distinct temperature relation of $~T^{-1.3}$. At first glance, these characteristics allude to a coherent, i.e. band-like charge carrier transport predominantly governed by inelastic scattering with accoustic phonons. However, as we will demonstrate, the experimental mobility data can be consistently described within the framework of incoherent, hopping-type transport modeled by Levich-Jortner rates, explicitly accounting for the inner and outer relaxation energies related to thermally induced lattice effects and enhanced electron-phonon interaction at elevated temperatures. Complementary band-structure calculations yielding temperature dependent effective mass tensors deviate stronger from experimentally observed spatially anisotropic transport behavior. Thus, these results hint at the fact that by the particular interplay of the transport energies the mobility of a given organic semiconducting material might appear to be band-like in a certain temperature regime even though the underlying charge carrier transport can be of incoherent, hopping-type nature. Building on this description, we further explore the role of the intermolecular electronic coupling and develop a procedure to distinguish between its dependence on static and dynamic lattice deformation upon temperature variation.  

\end{abstract}

\pacs{}

\keywords{}

\maketitle

%Introduction
\section{Introduction}
Charge carrier transport constitutes a decisive process in the operation of organic thin film devices such as light emitting diodes, photovoltaic cells and field-effect transistors (FETs).\cite{book:Bruetting2012} Despite tremendous efforts to develop high mobility organic semiconductor materials and their first commercial applications in displays and lightning, the microscopic mechanisms mediating the motion of electrons and holes on a molecular level and their interference with effects on spatially extendend length scales, in particular static or dynamic lattice deformations, have not yet been fully elaborated. A major challenge of up-to-date theoretical models is posed by the delicate energetic balance between the individual transport processes, demanding for reliable mobility data over a broad temperature range and along different crystallographic directions to cope also with the spatial anisotropy of charge carrier motion in crystalline molecular solids.\cite{inbook:Karl2001} In this respect, organic single crystals constitute reference systems due to their material-inherent electronic behavior with only minor influences by chemical or structural inhomogeneities.\cite{Pflaum2006} In combination with advanced simulations they enable detailed understanding of the charge carrier migration processes in relation to molecular packing, orientation and localization.\cite{Hannewald2004a,Troisi2007} A detailed overview dealing with the question of charge localization in different transport regimes has recently been provided by Fratini et. al.\cite{Fratini2016}

An elegant way to access these transport properties, that has recently been rendered successful, is based on the lamination of micrometer thick molecular crystals on vacuum-gap poly(dimethylsiloxane) (PDMS) substrates with predefined FET bottom-contact structures.\cite{Menard2004} To obtain the full information on the $in$-$plane$ charge carrier mobility this experimental approach requires, in principle, just one crystalline sample without need for destructive readjustment.\cite{Xie2013,Reese2007} But even though a band-like motion of electrons or holes might be anticipated and has been demonstrated for various polyaromatic single crystals e.g. by time-of-flight measurements,\cite{Warta1985,Tripathi2007} comparable FET studies on laminated crystals predominantly reveal a thermally activated transport characterized by a mobility decrease towards lower temperatures.\cite{Xia2007} This becomes even more remarkable considering the high $in$-$plane$ mobilities of > 10 cm$^{2}$/Vs observed already at room temperature in the case of laminated rubrene crystals.\cite{Sundar2004} The absence of a non-thermally activated charge carrier motion renders it difficult to judge whether the observed behavior is caused by extrinsic obstructions such as morphologically or chemically induced trap states at the crystal surface or by intrinsic limitations, namely the small band dispersion and strong phonon coupling expected for van-der-Waals bound solid states.\cite{Hannewald2004b} To discriminate the various contributions to charge carrier migration and to advance existing transport models it is therefore mandatory to have organic semiconductor references showing a non-dispersive transport together with spatially anisotropic effects along the various crystallographic directions.

We aimed for these challenges by investigating the temperature dependent charge carrier transport in \textit{N,N'}-Bis-(heptafluorobutyl)-2,6-dichloro-1,4,5,8-naphthalene tetracarboxylic diimide (Cl$_{2}$-NDI) single crystals. This particular naphthalene diimide derivative has not only become a potential candidate for complementary logical circuits by its auspicious high electron mobility of 1 cm$^{2}$/Vs in thin films \cite{Oh2010,Stolte2012} and up to 8.6 cm$^{2}$/Vs in solution processed micro-crystals,\cite{He2013} but can also be grown as single crystals by physical vapor deposition technique \cite{Laudise1998} under ambient conditions \cite{He2015} as utilized in this paper. The habit of the sublimation grown Cl$_{2}$-NDI crystals is characterized by lateral dimensions along the (001) plane of several millimetres and thicknesses of several tens of microns making elastomeric FET measurements of the anisotropic $in$-$plane$ mobility feasible. Advantageously, both laminated crystal and PDMS substrate are expected to exhibit similar thermal expansion coefficients thus maintaining the structural integrity of the samples by avoiding significant interfacial strain at the examined temperatures between 175 K to 300 K. We restricted our transport studies to this range to avoid interferences with a polymorph occurring at lower temperatures.\cite{He2015} Furthermore, operating the PDMS FET structures under vacuum as done in this study minimizes chemical surface reactions, which might lead to undesired effects by interfacial doping or dipole formation in the active channel region.\cite{Menard2004} Altogether, these conditions allow for reliable investigation of the electron mobility which act as a test-bed for theoretical calculations of this key parameter as a function of temperature and crystallographic direction.

\section{Experimental}
\label{Experimental}
Single crystals of the triclinic $\beta$-phase polymorph ($P \overline{1}$, $a$ = 5.346 \AA, $b$ = 6.317 \AA, $c$ = 19.157 \AA, $\alpha$ = 92.63 $^\circ$, $\beta$ = 98.55 $^{\circ}$\xspace, $\gamma$ = 109.00 $^{\circ}$\xspace at $T$ = 300 K; CCDC: 1472770)\cite{Note_Structure} were grown from purified Cl$_{2}$-NDI material by the physical vapor transport method \cite{Laudise1998} at $235 \pm 5$ $^{\circ}$C under ambient air.\cite{He2015} The transistor geometry utilized in this study consists of lithographically structured elastomeric PDMS supports covered by 3 nm thick Cr and 30 nm thick Au layers to form the contacts. By the elevated source and drain contacts a vacuum gap of about 4.9 $\mu$m in height and channel lengths of 50 $\mu$m are generated. To improve charge carrier injection the pre-structured Au contact pads have been surface functionalized by spin coating an aqueous solution of 0.004 wt.\% polyethylenimine ethoxylated (PEIE) onto the substrates.\cite{Zhou2012} All $I(V)$-measurements have been carried-out in transistor geometry (i.e. with two probe source and drain contacts) under high vacuum at a base pressure of $10^{-6}$ mbar in a Janis Research ST-500 micromanipulated probe station using an Agilent 4155C parameter analyzer. Electron mobilities were calculated in the linear regime of the transfer characteristics using Eq.~\ref{eq:mob}:
\begin{equation}
	\mu_{lin}(V_G)=\frac{\delta I_D}{\delta V_G}\frac{L}{WC_iV_D} \quad .
	\label{eq:mob}
\end{equation}

The temperature dependent crystallographic structure of the bulk samples was determined by a Bruker D8 Quest diffractometer, whereas the complementary assignment of the crystallographic axes was accomplished by a Bruker D8 Discover diffractometer, both using copper $K_{\alpha}$ radiation. 

\begin{figure}[t]
  \includegraphics*[scale=0.1]{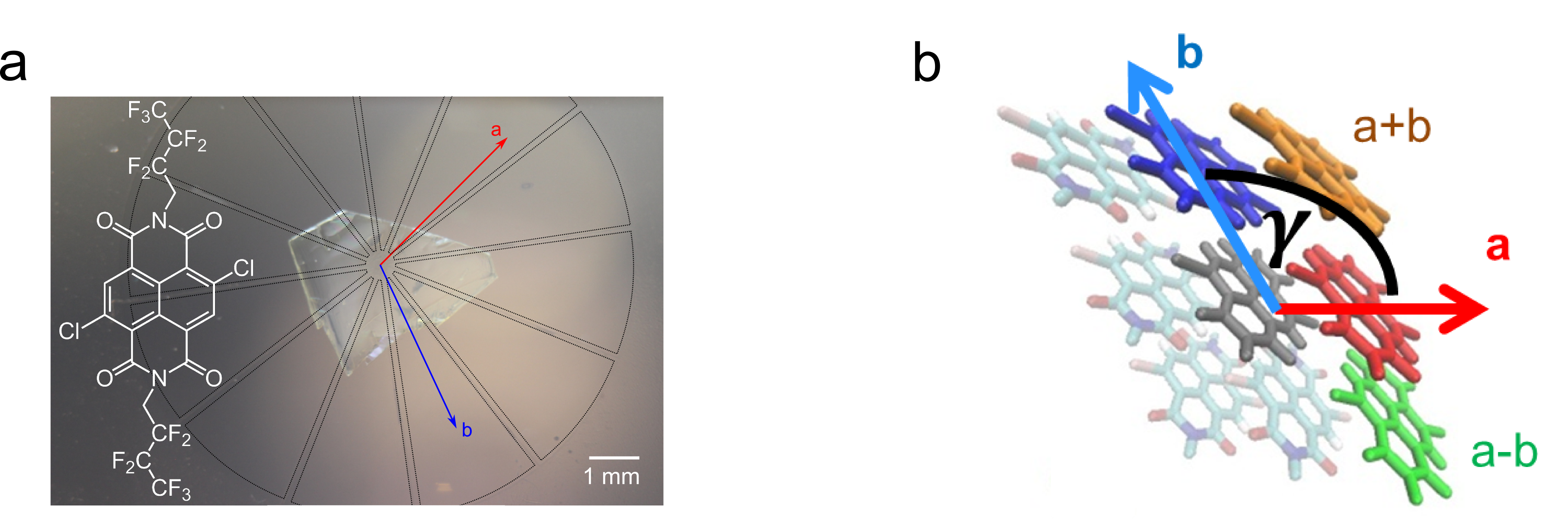}
  \caption{(a) Cl$_{2}$-NDI molecular structure and single crystal arrangement on top of an elastomeric FET support. (b) Molecular arrangement within the unit cell together with the relevant electron transfer directions $a$, $b$, $a+b$, $a-b$.}
  \label{fig:Transistor_picture}
\end{figure}

Figure~\ref{fig:Transistor_picture}(a) pictures the molecular structure of  Cl$_{2}$-NDI together with the alignment of a sublimation-grown crystal laminated on top of the pre-patterned PDMS support. The pie-shaped contact structure enables determination of the mobility along defined crystallographic directions at an angular resolution of 30 $^{\circ}$\xspace. The main charge transfer directions within the two-dimensional brickwall-type plane of slip-stacked Cl$_{2}$-NDI molecules, i.e. $a$, $b$, $a+b$ and $a-b$, are displayed in Figure~\ref{fig:Transistor_picture}(b).

%Simulation
\section{Theoretical decription by Levich-Jortner rates}
\label{sec:Simulation}
To provide theoretical access to the experimental transport data, charge carrier motion in organic semiconductors has to be divided into two limiting regimes depending on the degree of localization.%\newfalk{}{\cite{Ciuchi2012}} 
If the electronic coupling J between the molecules is weak, charge carriers are localized and transport can be described by non-adiabatic hopping rates depending on J$^2$.\cite{bredas_charge-transfer_2004}
 %Otherwise, adiabatic rates have to be used, which are independent of J.\cite{Yost2014}
In this case the time for the electronic wave function to move from donor to acceptor $\tau_\text{el}=\hbar/J$ (given by the reduced Planck constant $\hbar$ and the electronic coupling $J$) is much larger compared to the time for nuclear motion of the promoting mode $\tau_\text{nuc}=2\pi/\omega$. The high temperature limit of classical charge-transfer theory assumes that all promoting modes are harmonic and can be treated classically yielding the often used Marcus rate, where the activation energy is mainly given by the reorganization energy $\lambda$.\cite{marcus_electron_1993,hutchison_hopping_2005} Here, the latter is a sum of two contributions: a larger one caused by fast relaxation of nuclear coordinates of the two involved molecular species upon changing their charge state (intramolecular $\lambda^\text{int}$) and a smaller one originating from slow relaxation of the surrounding molecules (outer-sphere $\lambda^\text{out}$).

At ambient conditions, however, the fast intramolecular modes (typically C-C bond stretches with $\hbar\omega^\text{int} \approx$ 0.2 eV $\gg k_\text{B}T$) should be treated quantum-mechanically which is provided by the Levich-Jortner hopping rate,\cite{jortner_temperature_1976} where the activation energy is determined mainly by $\lambda^\text{out}$ 
\begin{eqnarray}
 k = &&\frac{2\pi}{\hbar}  \frac{|J|^2}{\sqrt{4\pi \lambda^\text{out} k_\text{B}T}} 
\sum_{N=0}^\infty \frac{S^N\exp (-S)}{N!} 
\exp
\left\{ -\frac{ \left[ \Delta E-\hbar N\omega^\text{int} -\lambda^\text{out} \right]^2}{4\lambda^\text{out} k_\text{B}T}
\right\} \,.
\label{eq:jortner}
\end{eqnarray}
Here it is assumed that the donor is initially in its vibrational ground state, while $N$ describes all accessible vibrational states of the acceptor at energy $\hbar\omega^\text{int}$, with $S=\lambda^\text{int}/\hbar\omega^\text{int}$ being the Huang-Rhys factor, $k_\text{B}T$ the thermal energy and $\Delta E$ the site-energy difference. 

While the hopping picture is appropriate for disordered amorphous organic semiconductors with large energetic disorder and low average electronic coupling $J\approx1$ meV, in organic crystals the strong $\pi$-$\pi$ interactions can lead to $J\approx\lambda^\text{out}\approx0.1$ eV rendering it $a priori$ not obvious whether a hopping model is applicable or not.
Therefore, we have compared the highest electronic coupling of J = 82 meV, which occurs along the $a$-direction in our Cl$_{2}$-NDI crystal at room temperature, with the following localization energies:

(i) intramolecular reorganization energy $\lambda^\text{int}$ = 370 meV, describing the relaxation of nuclear coordinates of a molecule upon (dis)charging in vacuum\cite{Troisi2007}, 

(ii) outer-sphere reorganization energy $\lambda^\text{out}$ = 40 meV, describing relaxation of the nuclei of surrounding molecules in the crystal after the charge-transfer process \cite{mcmahon_evaluation_2010,martinelli_influence_2010,norton_polarization_2008},

(iii) difference in polarization energy $\Delta$P = 300 meV  for an electron localized on a single molecule within the molecular crystal versus spreading the charge across a dimer, 

(iv) energetic disorder $\sigma$ = 40 meV, due to molecular vibrations associated with a distribution of electrostatic and polarization energies,

(v) intermolecular fluctuations (lattice phonons) leading to a broad distribution of J with a median of $<J>$ = 80 meV and a standard deviation of 31 meV along the a-direction at room temperature (10\% of molecular pairs have J < 50 meV and for 2\% of the pairs J < 30 meV).

Concerning point (v), it has been shown that the broadening of the J-distributions due to enhanced intermolecular fluctuations at elevated temperatures might lead to stronger charge localization, which in case of rubrene single crystals can explain the decrease in mobility with increasing temperature.\cite{Troisi2006,Fratini2016} However, the ratio between intramolecular reorganization energy and average transfer integral at room temperature along the direction of highest mobility amounts to $\lambda^\text{int}$ / $<J>$ = 370 meV / 80 meV = 4.6 in our Cl$_{2}$-NDI crystals, whereas for rubrene $\lambda^\text{int}$ / $<J>$ = 159 meV / 142 meV = 1.1. This results in an $a$ $priori$ stronger localization of charge carriers in our system and thus favors its description by a hopping rate model.

J and $\lambda^\text{int}$ were calculated by density functional theory (DFT). All DFT calculations in this work were performed using the B3LYP functional and the def-TZVP basis set in the TURBOMOLE package\cite{turbomole_2010_2} except for J where the functional was changed to BP86 for performance reasons.

The energy contributions (ii)-(iv) were obtained by the Thole model\cite{ruhle_microscopic_2011} based on atomistic charges. For the dimer configuration in (iii) we have equally placed $1/2$ of the anionic atomic charges on each of the two molecules and then used the Thole model to evaluate the polarization response of the surrounding molecular crystal. 

%According to their respective magnitude, $\lambda^\text{int}$ and $\Delta$P significantly outbalance the highest electronic coupling contribution and, since several other localization energies are of similar size as the electronic coupling, we considered the model of localized charges to be a valid approach for the temperature range investigated in our experimental transport studies.
Due to their fast time-scales, polarization interactions and intramolecular reorganization renormalize the transfer integrals $J$ to smaller values $J'$. It has been shown that for pentacene polarization leads to $J'/J = 0.64$.\cite{Bussac2004} With $\lambda^\text{int}$ = 96 meV and a mean vibrational energy of $\hbar\omega = 0.17 eV$ for this compound, the intramolecular reorganization $J'/J = exp[-\lambda^\text{int}/(2\hbar\omega)]$ yields an additional reduction by 0.75.\cite{Houili2006} The intramolecular contribution has been already accounted for in the Levich-Jortner model by the Huang-Rhys factor (eq. \ref{eq:jortner}). As in \cite{Bussac2004} the polarization effect has been estimated isotropic, this polarization effect has been neglected in our simulations of the anisotropic mobility tensor.
 
If we assume a similar effect of polarization in our Cl$_{2}$-NDI samples but consider the larger $\lambda^\text{int}$ = 370 meV at the same mean vibrational frequency, we arrive at a renormalized average coupling along the direction of highest mobility at room-temperature of $J'$ = 80 meV * 0.64 * 0.33 = 17 meV.
By comparing this value to the energetic disorder of $\sigma$ = 40 meV  the localization length of the charge carriers in units of intermolecular spacing L/a can be computed according to $L/a = 8(\sigma/J')^{-1.4}$,\cite{Houili2006} which finally results in a localization length of 2.4 molecular units. Since already the non-normalized integrals represent broad distributions and along the b-direction the integrals are even smaller, the tendency to localize is even further enhanced. This leads us to the conclusion that in our Cl$_{2}$-NDI molecular crystals a localized picture and thus a hopping transport mechanism are valid. 

 In order to complement this picture, we finally compare our temperature-dependent mobility-tensors based on the Levich-Jortner hopping rate to effective mass-tensors from a perdiodic DFT calculation of the band-structure (section \ref{sec:band}).

%Molecular dynamics
\subsection{Molecular dynamics}
\label{subsec:MD}
\begin{figure}[t]
  \includegraphics*[scale=0.076]{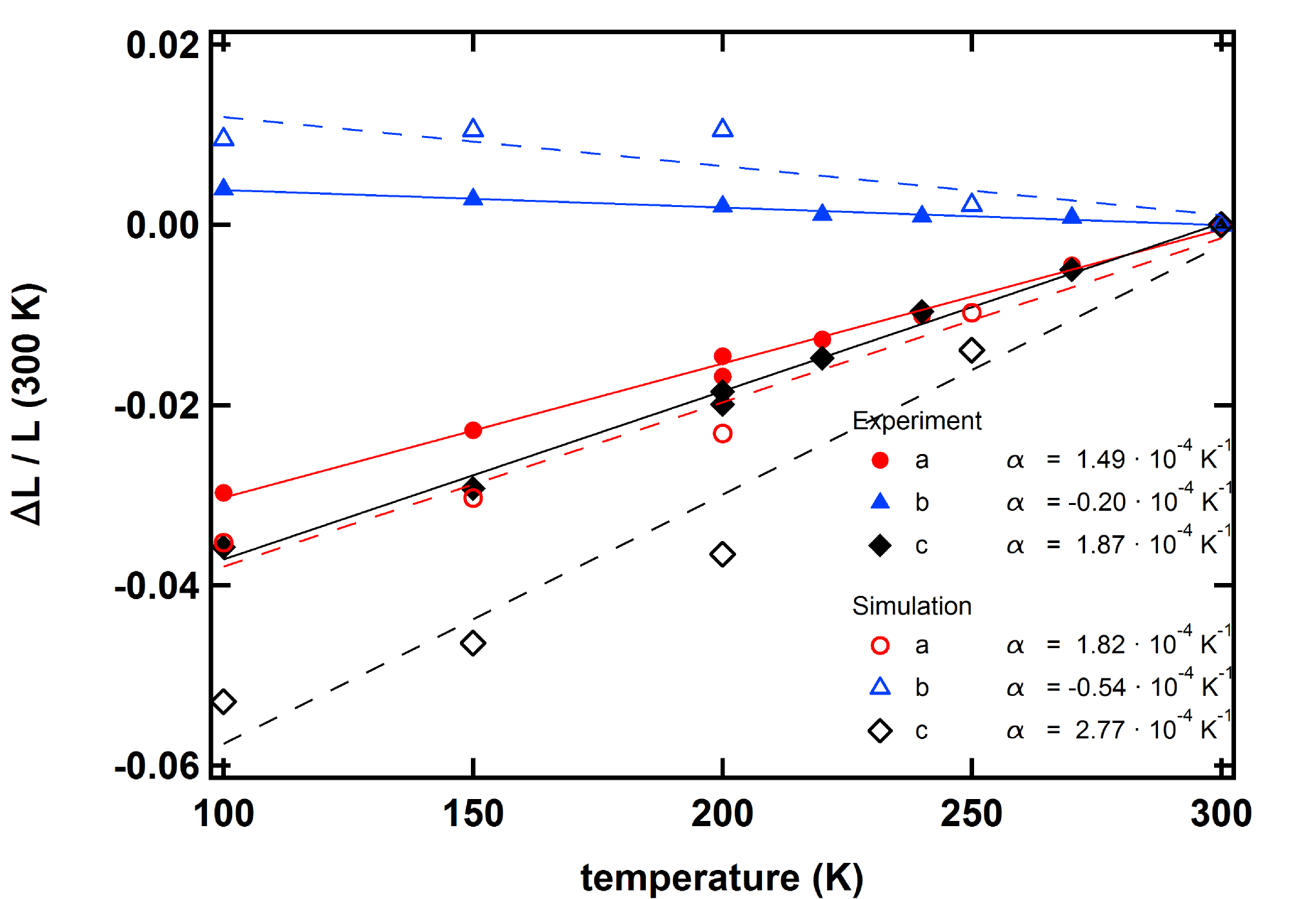}
  \caption{Relative lattice expansion $\Delta$$L$/$L(300K)$ along the crystallographic axes determined by X-ray analyses (full symbols) and force field simulations (open symbols) together with the respective thermal expansion coefficients, indicating the good agreement between theory and experiment within the examined temperature range.}
  \label{fig:thermal_expansion}
\end{figure}

We have used molecular dynamics (MD) simulations in order to study how temperature affects the morphology of the Cl$_{2}$-NDI crystal and thereby the electronic coupling between the molecules in the (001) plane.
A force field based on Optimized Potentials for Liquid Simulations (OPLS) parameters \cite{Watkins2001} was used with atomistic partial charges obtained from DFT while unknown intramolecular potential energy surfaces have been parametrized again using DFT.\cite{ruhle_microscopic_2011} 
The only modification to the original OPLS parameters was a reduction of the van der Waals radius of the Cl atoms by 12\%. 
When starting the MD simulation from the experimentally known crystal structure at $T=200$ K the equilibrated unit cell deviates by less than 1\% in all six lattice parameters ($a$,$b$,$c$,$\alpha$,$\beta$,$\gamma$) and we are also able to reproduce the experimentally observed thermal expansion of the crystal as shown in Figure~\ref{fig:thermal_expansion}.  As mentioned before, in this temperature range the crystal structure of sublimation grown Cl$_{2}$-NDI single crystals retain its triclinic symmetry \cite{He2015} and thus differs from the monoclinic structure previously reported for solution grown crystals.\cite{He2013} As indicated by Figure~\ref{fig:thermal_expansion} the experimental data for the thermal expansion coefficients, estimated by single crystal X-ray diffraction, can be well reproduced by our simulations based on molecular dynamics.

%Details on all MD simulations performed by the GROMACS\xspace package:
MD simulations were performed using GROMACS \cite{hess_gromacs_2008} for boxes containing 18x18x6 Cl$_{2}$-NDI molecules with periodic boundary conditions.
We simulated at constant particle number, pressure and temperature (NPT ensemble) with anisotropic Berendsen pressure coupling \cite{berendsen_molecular_1984} (reference pressure 1 bar, time constant of $\tau_P=5$ ps and compressibility $4.5\times10^{-5}$ bar $^{-1}$) in combination with stochastic velocity rescaling for temperature coupling \cite{bussi_canonical_2007} ($\tau_T=0.07$ ps) and a time step of 0.002 ps, using constrained bonds and the grid-based particle mesh Ewald technique for electrostatic interactions.\cite{perera_smooth_1995}
Morphologies for lower and higher temperatures were achieved by thermal annealing starting from the equilibrated MD structure at $T=200$ K with maximum rates of 25 K/ns.
By this protocol the deviation between equilibrated simulation and experiment for the most relevant lattice parameters within the transport plane $a$,$b$,$\gamma$ is less than 1\% for $T$=175K and $T$=300K.

\subsection{Temperature dependent transport parameters}
\label{sec:charge_transport}
In order to separate the different thermal effects on the anisotropic charge carrier transport we have, at first, utilized the lattice parameters from the MD-equilibrated crystal structure for $T$=300 K and $T$=175 K,
%this is an important point! While For the periodic Band-structure calculation we have utilized the experimental crystal structures at $T$=300 K (corrected for the side-chain) and at $T$=200 K
%we have used the lattice parameters from the equilibrated crystal structure for 200K and 300K obtained from the MD simulation for calculation of J for the Jortner rates!
keeping all molecules exactly at their lattice sites and with orientations defined by the crystallographic angles excluding any disorder ("frozen crystal"). To model the temperature and direction-dependent mobility we employ Eq.~\ref{eq:jortner}, where we assume $\hbar\omega^\text{int}=0.2$ eV and compute all parameters for electron transport by $ab-initio$ methods \cite{ruhle_microscopic_2011} yielding $\lambda^\text{int}=0.37$ eV (independent of temperature and direction) while all other transport parameters are summarized in Table~\ref{table:dips}.

In the frozen crystal (labeled cry in Table~\ref{table:dips}) the transport parameters are identical for all pairs that belong to the same crystallographic direction shown in Figure~\ref{fig:Transistor_picture}(b). 
The small intermolecular distance $d$ along the $a$-axis leads to the strongest electronic coupling $J$ and lowest $\lambda^\text{out}$ both favoring transport along this direction. 
Due to the nodal structure of the LUMO (lowest unoccupied molecular orbital) the coupling along $a+b$ is negligible although $d$ is not too large indicating that charge transport in this direction has to occur via sequential hopping in the $a$ and $b$-direction. The coupling along the $b$- and $a-b$-direction is slightly weaker compared to that along $a$ which effectively yields a two-dimensional transport within the brickwall-type packed Cl$_{2}$-NDI (001) plane. As the LUMO is located on the conjugated core of the molecule, transport along the $c$-direction, i.e. along the side chains, is neglected in our simulation. The biggest difference for the two principal transport directions $a$ and $b$ of the frozen crystal at 300 K and 175 K is the enlarged $J$ along the $a$-direction due to the anisotropic contraction of the crystal upon cooling.

As we have only one molecule per unit cell the energy difference vanishes, i.e. $\Delta E=0$.  
Furthermore $\lambda^\text{out} < \lambda^\text{int}$ and $\lambda^\text{out}$ decreases with intermolecular distance $d$.\cite{ruhle_microscopic_2011} Note that $\lambda^\text{out}<\hbar\omega^\text{int}$ as we have used a Pekar-factor of $c_p=0.05$ \cite{ruhle_microscopic_2011} (which linearily scales $\lambda^\text{out}$) such that only the first vibrational states contribute to Eq.~\ref{eq:jortner}.

\begin{table}
\begin{ruledtabular}
\begin{tabular}{l | l| lll | llll | lll | lll }
direction & $T$ [K]      & &$d$ [\AA]&             & &$J$ [meV]  & &                                  &  &$\Delta E$ &[meV]                & &$\lambda^\text{out}$ &[meV] \\ 
\hline
   &            &cry $\leftrightarrow$& <$\cdot$> &$\sigma$     &cry & $\lg[J^2]$ $\leftrightarrow$& <$\lg[J^2]$> &$\sigma$     &cry $\leftrightarrow$ &<$\cdot$>&$\sigma$    &cry $\leftrightarrow$ &<$\cdot$>&$\sigma$     \\
\hline
$a$ &300       	   & 5.35 & 5.36 & 1.71             &82 &-2.16   &-2.14&0.37              &0  &0&38      &41 &41&2       \\
    &175        	 & 5.20 & 5.20& 1.12              &102 &-1.99   &-2.03  &0.21         	&0  &0 &29      &38 &38 &2       \\
\hline

$b$   &300       	 & 6.31 & 6.30& 1.85              &58 &-2.47   &-2.43  &0.80             &0  &0 &35         &52 &51 &3       \\
   &175        	   & 6.38 & 6.38& 1.27              &57 &-2.48   &-2.18  &0.39          		&0  &0 &25       &53 &51 &2       \\
\hline

$a-b$   &300       & 9.45 & 9.45& 1.53              &16 &-3.55   &-3.70  &0.43                &0  &0 &40         &91 &90 &2       \\
   &175        	   & 9.40 & 9.41& 1.03              &18 &-3.49   &-3.56  &0.24          			&0  &0 &30         &79 &78 &2       \\
\hline

$a+b$   &300       & 6.90 & 6.90& 1.45              &0.2 &-7.51   &-7.16  &0.96                &0  &0 &39         &81 &80 &2       \\
   &175        	   & 6.87 & 6.85& 1.03              &0.1 &-8.05  &-7.30  &0.48          		  &0  &0 &29         &88 &89 &2       \\

\end{tabular}
\end{ruledtabular}
\caption{Transport parameters for Eq.~\ref{eq:jortner} in the frozen crystal (cry) as compared to median values <$\cdot$> and standard deviations $\sigma$ for the distributions induced by lattice phonons along crystal directions defined in Figure~\ref{fig:Transistor_picture}(b) for different temperatures.
Distributions of center of mass distance $d$, site-energy difference $\Delta E$ and outer sphere reorganization energy $\lambda^\text{out}$ are Gaussian. As electronic couplings $J$ are very broad and non-Gaussian we display the median and $\sigma$ for $\log_{10}[($J/eV$)^2]$. Thus less negative numbers correspond to higher $J$. For comparison we have also converted $J$ for the frozen crystal into $\log_{10}[($J/eV$)^2]$ with largest deviations from the frozen crystal occurring at 175 K along the b-direction.}
\label{table:dips}
\end{table}

The directional mobility $\mu=(\vec{v}\cdot\vec{F})/F^2$ at an angle $\phi$ to the $a$-axis can now be obtained from all hopping rates performing in average $100$ Monte-Carlo simulations for a single, randomly injected electron in a periodic box projecting its mean velocity $\vec{v}$ on the applied field of $F=10^7$ V/m.
%\newnhh{ The resulting mobility tensor for the frozen crystal is shown in Figure~\ref{fig:mob_novib}.
%For $T$ = 300 K the maximum mobility of $\mu_\text{max}=7.4$ cm$^{2}$/Vs is obtained at $\phi$ = 342 $^{\circ}$\xspace.
%When cooling to $T$ = 175 K the tensor rotates by 8 $^{\circ}$\xspace towards the $a$-axis and the overall mobility increases. The increase to $\mu_\text{max}=17.6$ cm$^{2}$/Vs by a factor of 2.4 can only partly be explained by a 1.5-fold increase of the rate-relevant $J^2_a$ along the $a$-axis upon cooling. The missing contribution can be deduced from the temperature dependence of Eq.~\ref{eq:jortner}.}{}

For a one-dimensional transport along an arbitrary direction ($z$) Levich-Jortner-hopping along ($k_+$) or against ($k_-$) a small electric field relates to a small energy difference 
$\Delta E=e F \Delta z \rightarrow 0$.
Implying that only $N$=0 significantly contributes in  Eq.~\ref{eq:jortner} yields the following temperature dependence of mobility:
\begin{equation}
\mu=\frac{\left(k_+-k_-\right)\Delta z}{F}=\frac{2\pi J^2 \exp(-S)}{\hbar\sqrt{{4\pi\lambda_\text{out} k_\text{B} T}}} \frac{e(\Delta z)^2}{k_\text{B} T}\exp\left(-\frac{\lambda^\text{out}}{4 k_\text{B} T}\right)\, \rightarrow \mu(T)\sim \frac{\exp\left(-\frac{\lambda^\text{out}}{4 k_\text{B} T}\right)}{T^{3/2}}\,.
\label{eq:mu_T}
\end{equation}
%
%\newnhh{This indicates that changing $T$ from 300 K to 175 K will increase $\mu$ along the $a$-direction by $f_a^T=1.7$ due to small $\lambda^\text{out}_a=0.04$ eV, while along the $a-b$-direction the increase is weaker $f_{a-b}^T=1.1$  due to larger $\lambda^\text{out}_{a-b}=0.1$ eV. Combining the gain in $J^2$ along the $a$-direction ($f^{J^2}_a$=1.5) with the gain due to temperature in the rate ($f_a^T=1.7$) explains the overall increase of $\mu_{max}$ by 2.4.}{}

Note that eq.~\ref{eq:mu_T} also can be deduced from the Marcus rate model, however, with $\lambda=\lambda^\text{int}+\lambda^\text{out} \approx 0.45$ eV in the exponent a {\it decrease} in mobility by more than one order of magnitude will occur upon cooling, contradicting the experimental findings (see Figure~\ref{fig:Tensor_exp_sim}(a) and discussion further below). In the case of Levich-Jortner hopping, $\lambda^\text{out}$ is small compared to $4k_{B}T$ (see Table~\ref{table:dips}). Therefore, the mobility is expected to follow the power law $\mu \sim T^{-3/2}$ for a temperature independent $J^2$. Note that this explanation of reduced mobility at higher temperature is different compared to the mechanism suggested in case of rubrene single crystals, where the increasing disorder due to stronger intermolecular fluctuations is assumed to decrease the localization length and thus the charge carrier mobility.\cite{Fratini2016}  But the predicted $T^{-2}$ dependence of the mobility is not supported by the experimentally observed $T^{-3/2}$ behavior of our samples. In contrast, the Levich-Jortner rate model can reproduce the measured $\mu(T)$ data and therefore, appears to be better suited.
%\textcolor{red}{Die J(T)-Tendenz wird in den folgenden Abschnitten besprochen, weil im letzten Abschnitt vor den conclusions sich darauf bezogen wird?} -->JA

Finally, we also account for lattice vibrations (phonons):
Our MD-simulation reveals that all molecules undergo vibrations around their lattice points which are well described by Gaussian distributions of center-of mass distances $d$ (median and standard deviation are labeled <$\cdot$> and $\sigma$ in Table~\ref{table:dips}, respectively). %
%Our force-field is well adapted as indicated by only a small shift of $d$ compared to the frozen crystal configuration. When cooling the crystal, phonons are frozen out leading to smaller $\sigma$.

As a consequence of these phonons each molecular pair acquires a unique conformation and environment which transform the anisotropic transport parameters into distributions. We have taken two representative snapshots, each containing 6x18x18 molecules, from our MD simulations after equilibration at 175 K and 300 K, respectively. Then all Jortner rates for our KMC simulations were evaluated by the parameters obtained from the molecular pairs within these snapshots.
Site energy differences $\Delta E$ computed including electrostatic and polarization interactions \cite{ruhle_microscopic_2011} are well described by Gaussian distributions (Table~\ref{table:dips}) with $\sigma_{\Delta E}$ representing the energetic disorder. As the energetic disorder is similar for all directions, its inclusion in Eq.~\ref{eq:mu_T} does not lead to a rotation or deformation of the tensor.
Molecular vibrations also result in Gaussian broadening of $\lambda^\text{out}$ due to different environments but their effect is weak as the standard deviation is small and isotropic.

While the dependence of electrostatic and polarization effects (both affecting $\Delta E$ and $\lambda^\text{out}$) on intermolecular distances is given by a power law, the electronic couplings $J$ depend exponentially on distance. 
Furthermore, via the nodal structure of the orbitals $J$ is also very sensitive to the relative orientation between two molecules. Therefore, the main influence on shape and anisotropy of the mobility tensor results from the temperature dependence of the $J$ distributions.
In case of Cl$_{2}$-NDI, just including distributions in $J$ on top of the static picture (while keeping $\Delta E=0$ and $\lambda^\text{out}$ fixed) does not lead to strong changes in absolute mobilities which is indicated by the small aspect ratio of $\mu$ around 3 (Figure \ref{fig:Tensor_exp_sim}b) allowing for effective 2 dimensional transport within the (001) plane. As a consequence, the few weak couplings in the low energy tail of $J$-distributions do not result in bottle-necks for charge transport.

%Band-structure details
\section{Band-structure calculations}
\label{sec:band}
Electronic-band-structure calculations for the two crystallographic structures obtained experimentally at 200 K and 300 K  were carried out using DFT and the Perdew Burke Ernzerhof Generalized Gradient Approximation GGA-PBE \cite{Perdew1996,Perdew1996_erratum} as implemented in Atomistix Tool Kit (ATK).\cite{AtomistixToolkit2014_2} The k-point sampling was 6x6x2 for both structures, and we set 75 Hartree as the cutoff energy and select 10$^{-5}$ Hartree as the convergence criterion for the total energy. The electron temperature is taken as 300 K. Upon obtaining the electronic-band-structure for both structures, the fractional k-point X(0.5, 0.0, 0.0) was identified as a conduction band minimum. For the calculation of the effective mass tensor and the effective mass direction, ATK's effective mass tool as implemented in version 2014.2 was used.

%200K (086_OFET1.cif)
%300K (130u_absnew_falk_corrected_sidechains_bym22_200K.cif)

%Results
\section{Experimental mobility tensor}
\begin{figure}[t]
  \includegraphics*[scale=0.1]{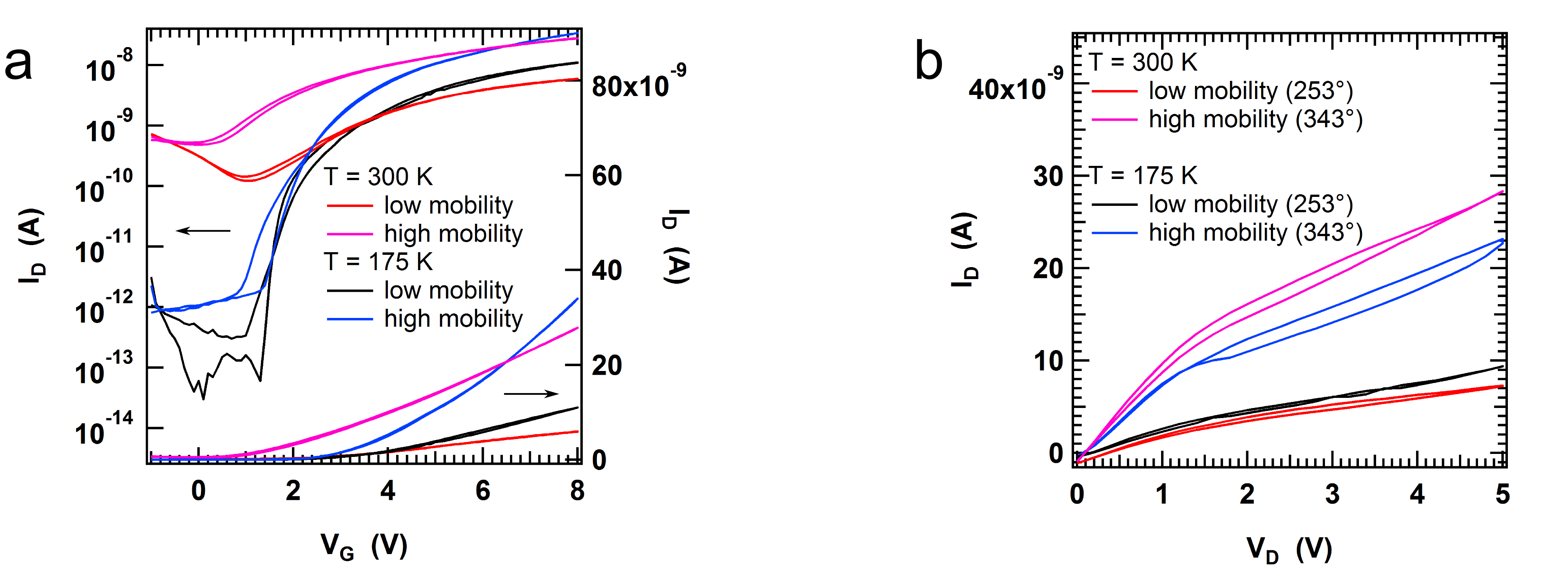}
  \caption{(a) Transfer characteristics measured at 175 K and 300 K along 253 $^{\circ}$\xspace (low mobility) and 343 $^{\circ}$\xspace (high mobility) with respect to the $a$-axis at $V_{\textrm{D}}$ = 1.5 V, and (b) corresponding output characteristics measured at $V_{\textrm{G}}$ = 5 V.}
  \label{fig:Transistor_IV}
\end{figure}

Performing the electrical characterization in FET geometry and in the temperature range described above, Figures~\ref{fig:Transistor_IV}(a,b) show representative transfer and output characteristics of a Cl$_{2}$-NDI single crystal FET recorded at 175 K and 300 K along the directions of low (253 $^{\circ}$\xspace) and high (343 $^{\circ}$\xspace) mobility (angles are always declared with respect to the $a$-axis). As indicated by the small hystereses in the $I(V)$-curves upon up- and down-cycling, dielectric relaxation losses caused e.g. by trap-assisted charge carrier localization or pronounced injection barriers are of negligible relevance for the deduced transport parameters.     
For single transistors, electron mobilities estimated in the linear transfer regime at 175 K by means of Eq.~\ref{eq:mob} range between 0.3 cm$^{2}$/Vs to 2.8 cm$^{2}$/Vs depending on the respective crystallographic direction. The well-defined On$/$Off states in our single crystal FETs are confirmed by their related current ratio of about $10^{5}$ which can be attributed to the small leakage currents of the utilized vacuum gap structure. Threshold voltages of about 3.5 $\pm$ 0.6 V indicate the presence of trap states which can be effectively filled already at a low density of accumulated charges considering the small gate-voltage and low capacitance of 0.18 nF/cm$^{2}$. Accordingly, at 300 K mean threshold voltages are reduced to 2.1 $\pm$ 0.6 V mainly due to thermal detrapping of electrons. The activation of otherwise immobilized charges at elevated temperatures is confirmed by enhanced leakage currents at room temperature leading to On$/$Off ratios of about ten and by a superimposed ohmic contribution in the output characteristics. Most strikingly, at 300 K the lowest and the highest $in$-$plane$ mobility are reduced to 0.1 cm$^{2}$/Vs and 1.5 cm$^{2}$/Vs, respectively.

\begin{figure}[t]
	\includegraphics*[scale=0.0570]{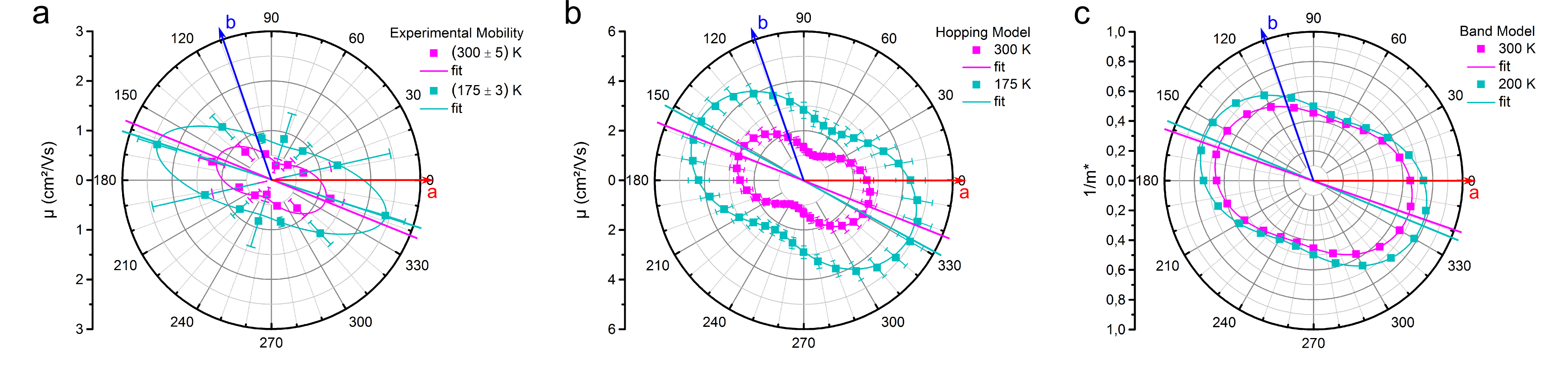}
  \caption{(a) Electron mobility tensors of Cl$_{2}$-NDI single crystals measured at 300 K (magenta) and 175 K (cyan). (b) The corresponding simulated tensors are calculated by a hopping-like transport model based on Levich-Jortner rates. The simulation includes static as well as dynamic lattice effects (see below: section~\ref{sec:static_dynamic}). (c) Effective mass-tensors from periodic DFT band-structure calculations based on experimental crystal structures at 300 K and 200 K.}
  \label{fig:Tensor_exp_sim}
\end{figure}

Measuring the FET characteristics along various directions in the Cl$_{2}$-NDI (001) plane the electron mobility tensor has been constructed. Exemplarily, the anisotropic electron mobility at 300 K and 175 K is displayed in Figure~\ref{fig:Tensor_exp_sim}(a). To estimate the range of confidence, mobility values along different directions have been averaged in compliance with the $P \overline{1}$ point group symmetry of the Cl$_{2}$-NDI $\beta$-phase crystal structure (see above), with deviations visualized by the error bars. At room temperature the angular dependent mobility resembles a constricted ellipse with its long axis tilted by -22 $^{\circ}$\xspace and its short axis by -112 $^{\circ}$\xspace with respect to the $a$-axis. The mobility aspect ratio along the principal directions $\mu_{\textrm{max}}/\mu_{\textrm{min}}$ amounts to three and thus resembles that of other organic compounds like Pentacene,\cite{Lee2006} Rubrene\cite{Reese2007,Sundar2004} or DCNQ.\cite{Menard2004} 
Upon cooling to 175 K, the overall electron mobility increases by a factor of two with the principal axes of the mobility tensor remaining fixed within the angular resolution of $\pm$ 5 $^{\circ}$\xspace.

\begin{table}
\begin{ruledtabular}
\begin{tabular}{ l | ccc | ccc | ccc }
 & &300 K & & 175 K & 175 K & 200 K & &Change&\\ 
\hline
 & Exp & Hopping & Band & Exp & Hopping & Band & Exp & Hopping & Band \\
\hline
$\overline{\mu}_{\textrm{max}}$ (cm$^2$/Vs) & 1.2 & 2.8 & & 2.4 & 4.8 & & x 2.0 & x 1.7 & \\
$\phi_{\textrm{max}}$ ($^{\circ}$\xspace) & 158 / 338 & 158 / 338 & 161 / 341 & 162 / 342 & 152 / 332 & 158/ 338 & + 4 & - 6 & - 3\\
$\overline{\mu}_{\textrm{min}}$ (cm$^2$/Vs) & 0.4 & 1.1 & & 0.7 & 2.3 & & x 1.8 & x 2.1 & \\
$\phi_{\textrm{min}}$ ($^{\circ}$\xspace) & 68 / 248 & 68 / 248 & 71 / 251 & 72 / 252 & 62 / 242 & 68 / 248 & + 4 & -6 & - 3\\
$\overline{\mu}_{\textrm{max}}/\overline{\mu}_{\textrm{min}}$ & 3.0 & 2.5 & 1.6 & 3.4 & 2.1 & 1.7 & x 1.1 & x 0.8 & x 1.1\\

\end{tabular}
\end{ruledtabular}
\caption{Measured and calculated mean values of the lowest and highest mobility as well as of their angular relation with respect to the crystallographic $a$-direction at 300 K and 175 K (band: 200 K). The variations upon cooling are listed in right column and indicate the good agreement between experiment and theoretical predictions by the Levich-Jortner rate model except for some slight deviation in the tensor rotation at low temperatures.}
\label{table:Vergl_Mob}
\end{table}

%Simulated results
\section{Simulated mobility tensor}
\label{sec:Simulated mobility tensor}
Complementary to the experimental results, the anisotropic mobility tensor at room temperature has been calculated by the theoretical Levich-Jortner rate model introduced in detail in section~\ref{sec:Simulation}. The predicted electron mobilities as well as their spatial anisotropy are illustrated in Figure~\ref{fig:Tensor_exp_sim}(b) exhibiting a good agreement with the experimental data, in particular concerning the difference in the absolute mobility values by only a factor of two. At room temperature both experiment and simulation reveal the direction of highest mobility at an angle of 338 $^{\circ}$\xspace and display a pronounced constriction in the direction of lowest mobility.
At 175 K the overall mobility values increase by a factor of two leaving the aspect ratio $\mu_{\textrm{max}}/\mu_{\textrm{min}}$ almost unchanged, except for a partial lifting of the constriction along the direction of lowest mobility. For all of these transport characteristics, which are summarized in Table~\ref{table:Vergl_Mob}, the good agreement between the experimental data and the simulations corroborate the suitability of the Levich-Jortner rate model. Note that the use of Marcus rates would lead to decreasing mobilities with decreasing temperature without affecting the rotation of the tensor significantly. Therefore, the temperature dependent behavior of the electron mobility tensor establishes fundamental criteria to judge on suited transport models and thereby, on the essential microscopic charge migration mechanisms. Within the chosen approach we were able to predict these variations, which are mainly determined by J$^2$.

In order to complete the picture we have performed band-structure calculations by periodic DFT based on the experimental crystal structures at 300 K and 200 K (details are given in Section~\ref{sec:band}). From the band-structure we extracted the inverse effective electron mass-tensor, which is displayed in Figure~\ref{fig:Tensor_exp_sim}(c). In band-transport the mobility $\mu=e\tau/m^*$ depends on the electron charge $e$, the scattering time $\tau$ and the inverse effective mass $1/m^*$. Because $\tau$ can cover several orders of magnitude\cite{Burland1977} we decided to not calculate the mobility. Instead, one can compare the direction and shape of the inverse effective mass tensor to the mobility tensors in Figure~\ref{fig:Tensor_exp_sim}(a,b) if one assumes $\tau$ to be isotropic and temperature independent. At 300 K the direction of highest mobility is found at an angle of 341 $^{\circ}$\xspace, the direction of lowest mobility at 251 $^{\circ}$\xspace. In comparison to the simulations based on Levich-Jortner rates only a small difference of 3 $^{\circ}$\xspace is observed for both directions, which shows that the electronic coupling is related to the band structure. This match is to be expected as both models rely on J$^2$. Going towards 200 K the directions of highest and lowest mobility as well as the aspect ratio do not change significantly.

The biggest difference between the inverse mass tensor and the mobility tensor from the Jornter-rates is (i) a less-pronounced constriction along the $a+b$-direction especially at 300K and (ii) a lower aspect ratio $\mu_{max}/\mu_{min}$ in case of the former (see Table \ref{table:Vergl_Mob}).
As the mobility tensor deduced from the Jortner-rates qualitatively agrees better with the experimental findings in those two points we assume a localization picture to be justified.

However, despite the remarkable agreement between experimental and Levich-Jortner mobility values at 300 K and 175 K and their overall spatial anisotropy, a small but nevertheless important deviation can be realized by the rotation of the principal axes upon cooling which shed light on the underlying transport processes. While in the measurements the tensorial rotation at 175 K occurs to be small ($\leq 4$ $^{\circ}$\xspace towards the $a$-axis) and within the experimental error, the theoretical calculations yielded a rotation by about $6$ $^{\circ}$\xspace and, even more strikingly, away from the $a$-axis. As this effect appears to be of thermal origin, we will address in the following the two major contributions to the temperature dependent transport in organic single crystals, namely thermally-induced lattice distortions leading to a static variation in the LUMO overlap along the anisotropic intermolecular distances within the unit cell \cite{Li2012} and, secondly, the thermal occupation of phonon modes and their spatially anisotropic coupling to charge carrier motion.\cite{Hannewald2004a} 

%static/dynamic picture
\section{Static and dynamic picture}
\label{sec:static_dynamic}
\begin{figure}[t]
  \includegraphics*[scale=0.072]{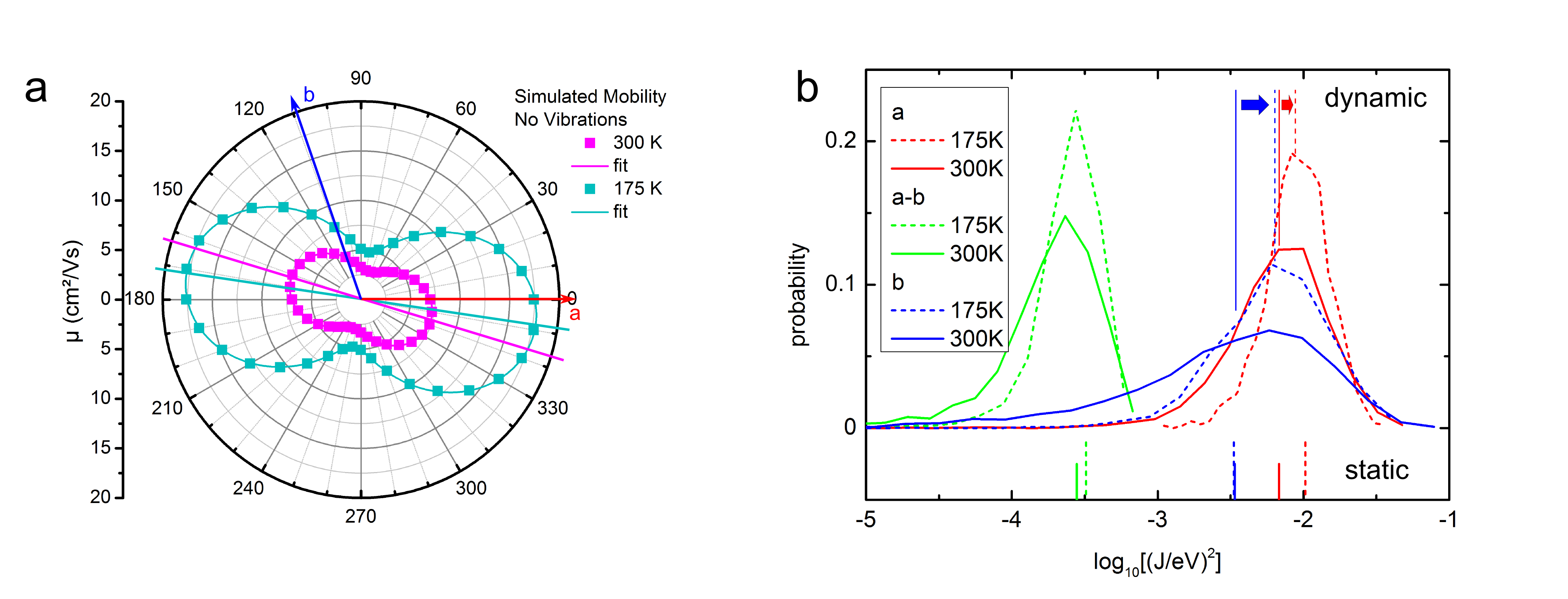}
  \caption{(a) Simulated mobility tensors using morphologies from experimental crystal structures at different temperature, however without molecular vibrations. (b) Distribution of electronic coupling elements in a static picture (lower part) and in presence of non-local phonon modes (upper part) together with their temperature dependent redistribution along different directions. While at 175 K the static picture shows improved coupling mainly along \textit{a}, including phonon modes also improves coupling along \textit{b} due to a redistribution of weight from the low-energy tail as indicated by the median shift of the distributions displayed by the upper lines and arrows.}
  \label{fig:mob_novib}
\end{figure}

To judge on the first contribution to the temperature dependent mobility, we investigated the thermally induced lattice expansion, i.e. the static picture, along the main crystallographic directions of the Cl$_{2}$-NDI unit cell between 100 K and 300 K as described in section~\ref{subsec:MD}. For this purpose, the well-known crystal structure at $T = 200$ K was taken as a starting point for further equilibration using a force field based on OPLS parameters.\cite{Watkins2001} The converging parameters of the unit cell coincide with the experimental ones within an error of 1 \%. From here, all other simulated crystal structures were obtained by thermal annealing. Disregarding fluctuations in inter-molecular distances and orientations (phonons) in the static picture (static LUMO overlap), we could investigate the structural influence on the charge transport separately. As key result, upon cooling from 300 K to 175 K the mobility tensor rotates by 8 $^{\circ}$\xspace towards the $a$-axis (see Figure~\ref{fig:mob_novib}(a)), which can be explained by the contraction of the $a$-axis and the simultaneous expansion of the $b$-axis upon cooling (see Figure~\ref{fig:thermal_expansion}), which results in an enhanced electronic coupling in the $a$-direction compared to the $b$-direction at 175 K, as can be seen in Figure~\ref{fig:mob_novib}(b). For this purpose, we have indicated the coupling elements of the static picture by lines in the lower part of this figure as compared to their distributions in presence of phonon modes above (which will be discussed below). We have omitted the $a+b$-direction as by the nodal structure of the LUMO in combination with the face-to-face orientation of molecular pairs the squared coupling is on average five orders of magnitude weaker than e.g. for the $a$-axis. Due to the contraction along the $a$-axis at low temperatures, the static picture results in improved coupling along the $a$-direction yielding a rotation of the mobility tensor towards this axis.

Mobilities are larger in the static case (Figure~\ref{fig:mob_novib}(a)) compared to the dynamic case of (Figure~\ref{fig:Tensor_exp_sim}(b)) due to energetic disorder which is missing in the former (see section~\ref{sec:charge_transport}). The increase of the mobility upon cooling by a factor of 2.4 to $\mu_\text{max}=17.6$ cm$^{2}$/Vs can only partly be explained by the 1.5-fold increase of the rate-relevant $J^2_a$ along the $a$-axis. The missing contribution can be deduced from the temperature dependence of Eq.~\ref{eq:mu_T}. This Equation indicates that changing $T$ from 300 K to 175 K will increase $\mu$ along the $a$-direction by $f_a^T=1.7$ due to small $\lambda^\text{out}_a=0.04$ eV, while along the $a-b$-direction the increase by $f_{a-b}^T=1.1$ is weaker due to larger $\lambda^\text{out}_{a-b}=0.1$ eV. Combining the gain in $J^\text{2}_{a}$ along the $a$-direction ($f^{J^2}_a$=1.5) (see Table~\ref{table:dips}) with the thermally induced gain in the rate ($f_a^T=1.7$) we are able to explain the overall increase of $\mu_{max}$ by 2.4. This significant increase along the $a$-direction compared to the $b$-direction and the resulting rotation of the mobility tensor in the static picture is not reflected in the experimental tensor. Therefore, the static lattice deformation alone is not sufficient for the description of the electronic coupling upon temperature variation.

%(Typ1: (P-1, a = 5.345 \AA, b = 6.317 \AA, c = 19.108 \AA, $\alpha$ = 82.10 \degree, $\beta$ = 82.49 \degree, $\gamma$ = 70.99 \degree)

Now we adress the second effect, the presence of phonon modes, i.e. thermally induced fluctuations of inter-molecular distances and orientations. They broaden the delta-like shape of the electronic coupling elements which have sufficed the static picture accounting solely for thermal expansion. Therefore, the activated phonon modes lead to a reduction of mobility (see Figure~\ref{fig:Tensor_exp_sim}(b)) by a factor of 2.6 for $T=300$ K and by 3.7 for $T=175$ K with respect to the mobility of the frozen crystal approach (Figure~\ref{fig:mob_novib}(a)). The reduction is stronger in case of lower temperature as $\hat{\sigma}=\sigma_{\Delta E}/(k_\text{B} T)=2.0$ and 1.6  for $T=175$ K and 300 K, respectively, and $\mu\sim\exp[-(C\hat{\sigma})^2]$ \cite{bassler_charge_1993} with a constant $C$ of the order of one.

The distributions of $J$ are non-Gaussian and very broad as shown logarithmically for the rate-relevant $J^2$ in Figure~\ref{fig:mob_novib}(b). For the principal transport directions $a$, $b$ and $a-b$, the biggest effect upon cooling occurs along $b$, where the distribution of $J$ is broadest: the median is shifted to much higher values and the low-energy tail is reduced significantly upon cooling although the mean of $d_b$ increases only slightly (and therefore $J_b$ in the frozen crystal does not change upon cooling). The final mobility tensors including all vibrational broadenings related to $J$, $\Delta E$, and $\lambda^\text{out}$ are displayed in Figure~\ref{fig:Tensor_exp_sim}(b) and show that the rotation towards the $a$-axis caused by thermal contraction of the crystal at lower temperatures (Figure~\ref{fig:mob_novib}(a)) is outbalanced by the changes in distributions of $J$ which leads to an overall rotation away from the a-axis.

%\newnhh{However, including phonon modes in addition to the thermal expansion, the anisotropic narrowing of the distributions upon cooling supports couplings along the $b$-direction even more than along the $a$-direction. While for the $b$-direction cooling increases the amount of couplings in the distribution above the static value from 50 \% to 75 \%, the amount of couplings that are higher than the static value along the $a$-direction is actually reduced from 50 \% to 40 \%. This redistribution of weight from the low-energy tail of the broadened coupling along the $b$-direction therefore outbalances the contribution from the neat static picture yielding an overall rotation of the tensor \textit{away} from the $a$-axis when thermal expansion \textit{and} phonon modes are both included at high and low temperature.}{}

Comparing the theoretical predictions on the temperature induced static effects alone (expansion of the lattice $\rightarrow$ rotation by 8 $^{\circ}$\xspace towards $a$-axis) versus those including also dynamic effects (expansion \textit{and} phonon modes $\rightarrow$ rotation by 6 $^{\circ}$\xspace away from the $a$-axis) one rationalizes that the two effects compete against each other with the resulting rotation of the tensorial mobility at low temperatures depending on their respective size and thus being characteristic for the organic material under study. In particular, for the anisotropic transport behavior observed in our Cl$_{2}$-NDI single crystals we identified almost a compensation of both effects leading to a spatially fixed alignment of the principal axes of the electron mobility tensor (Figure~\ref{fig:Tensor_exp_sim}(a)), whereas the simulations hint at a slight overestimation of the broadening of the electronic coupling distribution by phonons and its contribution to the electronic transport (Figure~\ref{fig:Tensor_exp_sim}(b)).

\section{Spatial anisotropy of temperature dependent mobility}
\begin{figure}[t]
  \includegraphics*[scale=0.095]{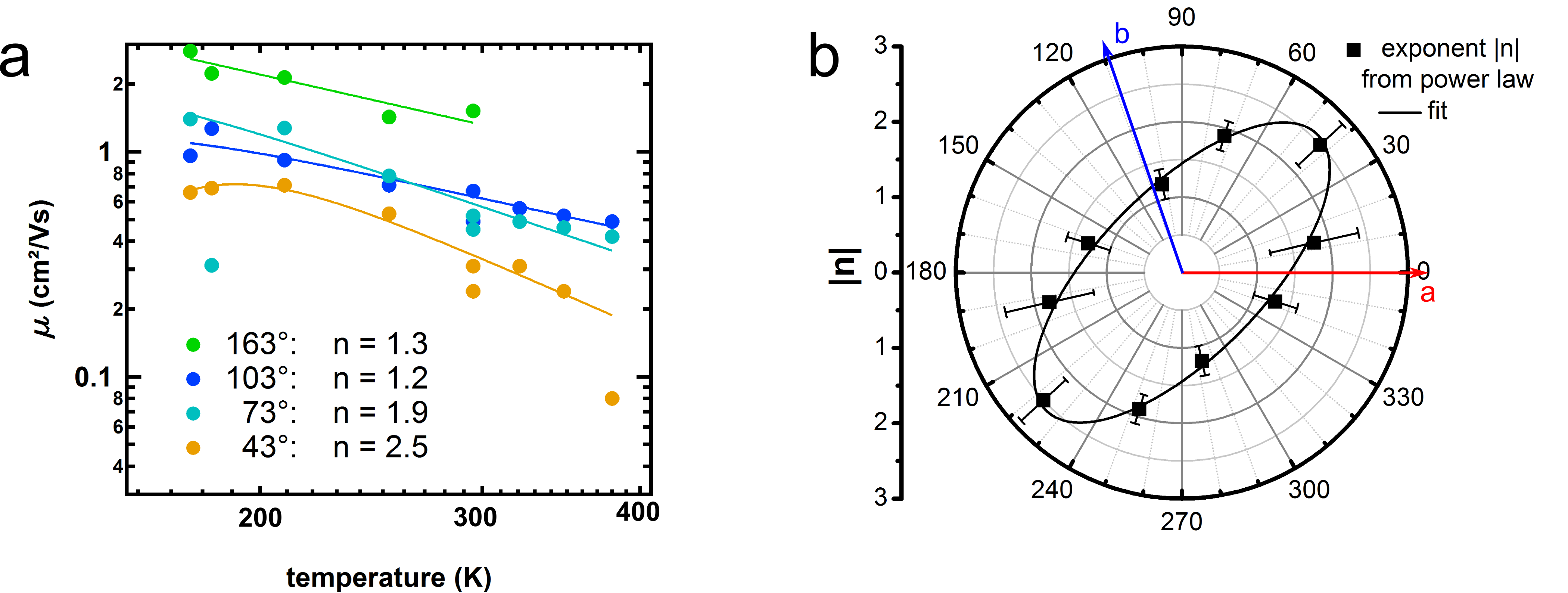}
  \caption{(a) Temperature dependent electron mobility measured along 163 $^{\circ}$\xspace, 103 $^{\circ}$\xspace, 73 $^{\circ}$\xspace and 43 $^{\circ}$\xspace with respect to the $a$-axis and fitted by the Hoesterey-Letson transport model. Below 200 K trap limited transport occurs only at 43 $^{\circ}$\xspace, i.e. close to the $a+b$ direction. (b) Absolute values of the power law exponent $n$ along different transport directions. To illustrate the entire angular behavior, half of the data points is mirrored according to the inversion symmetry of the Cl$_{2}$-NDI single crystals.}
  \label{fig:thermal_transport}
\end{figure}

Finally we adress the temperature dependent mobility along the principal crystallographic axes by fitting with the Hoesterey-Letson (HL) model.\cite{Hoesterey1963} This model implies the thermal release of charge carriers being localized by Boltzmann-distributed traps at an effective energy $E_{\textrm{tr}}$ and density $N_{\textrm{tr}}$. Even though the HL-model was introduced originally for oligoacene single crystals showing a band-like transport, the intrinsic mobility in this description, $\mu_0$, can be adapted to the respective conduction behavior, i.e. to either band- or hopping-type transport, with its temperature dependence described by a power law. This law typically accounts for the dominant scattering mechanism in band transport by its exponent $n$.\cite{Schein1977} In this case, we expect $n \approx$ 1.5 in eq.~\ref{eq:hoesterey}, not because of the scattering by acoustic phonons but for the temperature dependence of the Levich-Jortner model in eq.~\ref{eq:mu_T} for the case that $\lambda^\text{out}<4 k_\text{B} T$. The denominator limits the mobility at low temperatures due to increased trapping times
\begin{equation}
	\mu=\frac{\mu_0\cdot T^{-n}}{1+\frac{N_{tr}}{N_c}\cdot exp(\frac{E_{tr}}{kT})} \quad .
	\label{eq:hoesterey}
\end{equation}
Figure~\ref{fig:thermal_transport}(a) displays the measured temperature dependent electron mobility, which can be reasonably fitted by the described model and reveals only the transport along the direction 43 $^{\circ}$\xspace, i.e. close to the $a+b$-direction, to be limited at lower temperatures. The latter is caused  by either trap states or the poor electronic coupling along this direction (see above). The obtained temperature exponent $n$ shows a distinct relation to the crystallographic transport directions as indicated by its angular variation in Figure~\ref{fig:thermal_transport}(b). The value of $n \approx$ 1.5 along the $a$-direction and $n \approx$ 1.3 along the $b$-direction assorts well with the expected value of $n \approx$ 1.5 according to eq.~\ref{eq:mu_T}. The reduced value in the $b$-direction might stem from the increased $\lambda^{out}$ which weakens the temperature dependency of the mobility due to the exponential term in eq.~\ref{eq:mu_T}.
Along the $a+b$-direction the exponent $n$ is about twice as large. As the electronic coupling $J$ is orders of magnitudes lower between molecules along this direction (see Table~\ref{table:dips}) charge carriers preferentially perform two consecutive hops along the $a$- and $b$-direction, which might lead to the increased value of $n$.

%As the temperature dependence of the intermolecular spacing is considered to be a possible origin of the observed temperature dependent effects on the mobility, we conclude by the good agreement with the experimental data that our theoretical modelling adequately takes into account the relevant contributions by the transfer integrals to electron transport and its directionality. The calculated changes in the intermolecular transfer integrals with temperature are illustrated in Fig.~\ref{fig:thermal_expansion}\colorbox[rgb]{1,1,0}{(c,d)}. Obviously, the augmentation of the transfer integrals along the main crystallographic directions at lower temperatures leads not only at an increase of mobility but also to a rotation of the corresponding tensor. Therefore, the Marcus-type transport model in combination with the thermally induced static lattice deformation is able to correctly describe the thermal effects on the mobility. 

%Conclusions
\section{Conclusions}
We presented a combination of approaches on the temperature-dependent anisotropic electron mobility in Cl$_{2}$-NDI single-crystals by means of FET measurements and theoretical modeling of the microscopic transport via the Levich-Jortner rate model. These simulations demonstrated that the rotation of the mobility tensor with respect to the crystallographic orientation provides a criterion on the thermally induced contributions by static and dynamic lattice deformations. We have also compared the experimental results to band-transport from periodic DFT calculations and find a better agreement for the hopping approach. As a main result, we were able to show that hopping can be the dominating charge transfer mechanism despite the observed $~T^{-n}$ dependence of the mobility. Thus, our results indicate that for a given organic semiconducting material an increase of mobility at lower temperatures can also be explaind by an incoherent motion of charge carriers thoroughly taking into account static and dynamic lattice effects.

\begin{acknowledgments}
We thank the BMBF (project POLYTOS; FKZ: 13N10205) and the DFG (project PF 385/11-1) for financial support. F.W. and J.P. acknowledge the Bavarian State Ministry of Science, Research, and the Arts for generous support within the framework of the Collaborative Research Network \textit{Solar Technologies Go Hybrid}.

N.H. Hansen and F. May contributed equally to this work.
\end{acknowledgments}

\bibliography{literature}

\end{document}